\documentclass[final]{aipproc}

\layoutstyle{8x11single}

\usepackage{color}
\usepackage{amsmath,amssymb,mathrsfs}
\DeclareSymbolFont{symbols}{OMS}{cmsy}{m}{n}

 \def\a{{\alpha}}
 \def\b{{\beta}}

\def\z{{\zeta}}
 
 \def\bp{{\mathbf p}}
 \newcommand{\ben}{\begin{eqnarray}}
 \newcommand{\een}{\end{eqnarray}}
 \def\be{\begin{equation}}
 \def\ee#1{\label{#1}\end{equation}}

\begin{document}

\title{The Boltzmann Equation in Special and General Relativity}


\author{Gilberto M. Kremer}{
              address={Departamento de F\'{\i}sica, Universidade Federal do Paran\'a, Curitiba, Brazil}}



\begin{abstract}
Relativistic field  equations for a gas in special and general relativity  are determined from the Boltzmann equation. The constitutive equations are obtained from the Chapman-Enskog methodology applied to a relativistic model equation proposed by Anderson and Witting. Two applications in general relativity are considered: one refers to a gas in a homogeneous and isotropic Universe where irreversible processes are present during its evolution; in the other it is analyzed a gas under the influence of a spherically symmetrical non-rotating and uncharged source of the gravitational field.
\end{abstract}

\maketitle


 \section{Introduction}

We may say that the development of the relativistic kinetic theory goes back to 1911 when J\"uttner \cite{J1} succeeded to derive an equilibrium distribution function for a relativistic gas, which in the non-relativistic limiting case reduces to the Maxwellian distribution function. The relativistic  Bose-Einstein and Fermi-Dirac  distribution functions were also derived by him in 1928 \cite{J2}. At the beginning of the 1940s the covariant formulation of the Boltzmann equation was proposed by Lichnerowicz and Marrot \cite{LM}, while the determination of the transport coefficients from the Boltzmann equation by using the Chapman-Enskog method was obtained by Israel \cite{I} and Kelly \cite{K} in the 1960s. The general relativistic formulation of the Boltzmann equation goes back also to the sixties of the last century when Chernikov \cite{C} analyzed a relativistic gas in a gravitational field.

The aim of this work  is to analyze the Boltzmann equation within the framework of special and general relativity and to derive the corresponding macroscopic field equations.

First the Boltzmann equation in special relativity is analyzed and  the balance equations for the particle four-flow and energy-momentum tensor are obtained. As in the case of non-relativistic fluid dynamics these equations are useful to describe a relativistic fluid with five scalar fields of particle number density, four-velocity and temperature. The constitutive equations for the non-equilibrium pressure, pressure deviator and heat flux, as well as the corresponding transport coefficients of bulk and shear viscosities and thermal conductivity, are determined by using the Chapman-Enskog methodology applied to the Anderson and Witting model of the Boltzmann equation \cite{AW}. From the knowledge of the constitutive equations, the system of balance equations turns out to become a system of field equations for the basic fields and problems related to relativistic fluids can be analyzed.

Next, relativistic fluids in general relativity are analyzed within the framework of Boltzmann equation. In the first application a gas in a homogeneous and isotropic Universe is considered where irreversible processes are present during its evolution. The irreversible processes are characterized by a non-equilibrium pressure  which is determined from the Anderson and Witting model by using the Chapman-Enskog methodology.  Another application refers to a gas - which is not the source of the gravitational field - under the influence of a Schwarzschild metric, which is the solution of Einstein's field equations for  a  spherically symmetrical non-rotating and uncharged source of the gravitational field. For this case, the expression for the momentum density balance equation is obtained in the non-relativistic limiting case and for weak gravitational fields.

In this work  Greek indices run from  $\a=0,1,2,3$, while Latin indices from $i=1,2,3$.
 \section{Boltzmann equation in special relativity}

We consider a single non-degenerate relativistic gas in a Minkowski space with metric tensor $\eta_{\a\b}={\rm diag}(1,-1,-1,-1)$, where a particle of rest mass denoted by $m$  is characterized by the space-time coordinates $(x^\alpha)=(x^0=ct, {\bf x})$
and  by the momentum four-vector $(p^\alpha)=(p^0, {\bf p})$, where $c$ is the speed of light. The  momentum four-vector  has a constant length $mc$ which implies that  $p^0$ can be expressed in terms of $\bf p$ by  $p^0=\sqrt{\vert
{\bf p}\vert^2 +m^2c^2}$.

 The state of the relativistic gas -- in the phase space spanned by the space-time and momentum coordinates -- is characterized by the one-particle distribution function
$f(x^\alpha, p^\alpha)=f({\bf x}, {\bf p}, t)$, where   the quantity $f({\bf x, p},t)d^3{ x}\, d^3{ p}=f({\bf x, p},t)dx^1 dx^2 dx^3 dp^1 dp^2 dp^3$
 gives the number of particles at time $t$ in the volume element $d^3x$ about ${\bf x}$ and with momenta in a range $d^3p$ about $\bf p$.

In relativity we have to take into account the behavior of the quantities under Lorentz transformations.  The volume element
$d^3 x d^3 p$  is  an invariant under Lorentz transformations, i.e., $d^3x d^3p=d^3 x'd^3p'$. Furthermore, the number of particles in a volume element is  a scalar
invariant, since all observers will count the same number of particles. Hence, we may conclude that the one-particle distribution function $f({\bf x}, {\bf p}, t)$ is also a scalar invariant. It can be proved that the volume element $d^3p$ is not an invariant but only the ratio
${d^3p'/ p'_0}={d^3p/ p_0}$.

The evolution of the one-particle distribution function in the phase space is ruled by the  Boltzmann equation, which in the absence of
external forces reads
\be
p^{\alpha}{\partial f\over \partial x^\alpha}=p^0{\partial f\over \partial x^0}+p^{i}{\partial f\over \partial x^i}
=\int
\left(f_*'f'-f_*f\right) \,F\,\sigma\,d\Omega
{d^3p_*\over p_{*0}}.
\ee{2}
In the above equation we have introduced the abbreviations
$f_*'\equiv f({\bf x,p}_*',t),$ $f'\equiv f({\bf x,p'},t),$ $f_* \equiv f({\bf x,p}_*,t),$ $f \equiv f({\bf x,p},t),$, where $\bp$ and $\bp_*$ denote
 the momenta of two particles before a binary collision    and $\bp'$ and $\bp'_*$ are the corresponding momenta after collision. The pre and post collisional momentum four-vectors are connected by the energy-momentum conservation law $p^\a+p_*^\a=p^{\prime\a}+p_*^{\prime\a}$.  Furthermore, $F =\sqrt{(p^\alpha_*p_\alpha)^2-m^4c^4 }$ is the invariant flux, which in the non-relativistic limiting case it is proportional to the modulus of the relative velocity.   The differential cross-section and the element of solid angle that characterize a binary collision are denoted by $\sigma$ and $d\Omega$, respectively.

 The multiplication of the Boltzmann equation (\ref{2}) by an arbitrary function $\psi(p^\b)$ and integration of the resulting equation with respect to the invariant $d^3p/p_0$, yields
 the following general equation of transfer
\be
{\partial\over \partial x^\alpha}\int\psi p^{\alpha}f{d^3p\over p_0}
={1\over 4}\int\underline{(\psi+\psi_*-\psi'-\psi_*')} \,(f_*'f'-f_*f)\,F\,\sigma\,d\Omega{d^3p_*\over p_{*0}}{d^3p\over p_0},
\ee{4}
  which is only a function of the space-time coordinates. As in the non-relativistic case the right-hand side of the above equation is obtained by evoking the symmetry properties of the collision operator. When the underlined term in (\ref{4}) vanishes, $\psi$ becomes a summational invariant. The necessary and sufficient condition to be a summation invariant is that $\psi$ must be given by $\psi=A+B_{\alpha}p^{\alpha}$,
where $A$ is an arbitrary scalar and $B_\alpha$ an arbitrary four-vector that do not depend on the momentum four-vector $p^\alpha$.

The macroscopic description of a relativistic gas may be characterized by the two first moments of the distribution function, namely
\be
N^\a=c \int p^\a\,f\,{d^3p\over p_0}, \qquad T^{\a\b}= c \int p^\a\,p^\b\,f\,{d^3p\over p_0},
\ee{5}
which are known as the particle four-flow and the energy-momentum tensor, respectively. Their balance equations  are obtained from the transfer equation (\ref{4}) by choosing $\psi=c$ and $\psi=c p^\b$, yielding
\be
\partial_\a N^\a=0,\qquad \partial_\a T^{\a\b}=0,
\ee{6}
respectively.
They represent the conservation laws of the particle four-flow and of the energy-momentum tensor.

In the analysis of relativistic fluids it is usual to decompose the particle four-flow and the energy-momentum tensor in terms of  quantities that appear in non-relativistic fluid dynamics, like particle number density $n$, energy per particle $e$, hydrostatic pressure $p$, non-equilibrium pressure $\varpi$, pressure deviator $\mathbb{P}^{\a\b}$ (the traceless part of the pressure tensor) and heat flux $q^\a$. There exist two decompositions for the particle four-flow and energy-momentum tensor as functions of the above mentioned quantities and they are known as the decompositions of Eckart and of Landau-Lifshitz (see e.g. \cite{1}). Here we shall adopt the Landau-Lifshitz decomposition, namely
\be
N^{\alpha}=nU^{\alpha} -{q^\alpha\over h },\qquad
T^{\alpha\beta}=\mathbb{P}^{\a\b}-\left(p+\varpi\right)
\Delta^{\alpha\beta} +{en\over c^2}U^{\alpha} U^{\beta}.
\ee{7}
The above decomposition  is written in terms of the four-velocity $U^\a$ (such that $U^\a U_\a=c^2$) and of the projector $\Delta_{\a\b}=\eta_{\a\b}-U_\a U_\b/c^2$ (such that $\Delta^{\a\b}U_\b=0$). Furthermore, we have introduced the enthalpy per particle $h=e+p/n$.

As in the non-relativistic case, the equilibrium is characterized by the condition $f_*'f'=f_*f$ which renders a vanishing collision term of the Boltzmann equation (\ref{2}). This last relationship implies that $\ln f'_*+\ln f_*=\ln f_*+\ln f$ and $\ln f$ represents a summational invariant whose expression is given by $\ln f=A+B_{\alpha}p^{\alpha}$. The terms $A$ and $B_\a$ are determined from the equilibrium values of the particle four-flow and energy-momentum tensor and it follows that the equilibrium distribution function is represented by the Maxwell-J\"uttner distribution function
\be
f^{(0)}={n\over 4\pi m^2c k T K_2(\zeta)}
e^{-{U^\alpha p_\alpha/k T}},\qquad \hbox{with}\qquad K_n(\zeta)=\left(\frac{\zeta}{2}\right)^n\frac{\Gamma(1/2)}{\Gamma(n+1/2)}\int_{1}^\infty e^{-\zeta y}\left(y^2-1\right)^{n-1/2}\,dy,
\ee{8}
denoting the modified Bessel function of second kind. Note that the distribution  function depends on $\zeta=mc^2/kT$ which represents the ratio of the rest energy $mc^2$ of a particle and the thermal energy of the gas $kT$, where $k$ is the Boltzmann constant and $T$ the temperature of the gas. The limiting case $\zeta\gg1$  represents a non-relativistic regime of the gas, while $\zeta\ll1$ its ultra-relativistic regime. For example, for an electron of rest mass $m\approx 9.1\times 10^{-31}$ kg we have that $\zeta=5.9\times10^9/T$, indicating that only for very high temperatures the electron might be considered in the relativistic regime.

If we are interested in the description of the relativistic gas as a five-field scalar theory characterized by the fields of particle number density $n$, four-velocity $U^\a$ and temperature $T$, we have to determine the constitutive equations for the  energy per particle $e$, hydrostatic pressure $p$, non-equilibrium pressure  $\varpi$, pressure deviator $\mathbb{P}^{\a\b}$ and heat flux $q^\a$ as functions of $(n, U^\a, T)$. The determination of the constitutive equations for the energy per particle and  hydrostatic pressure follows from the insertion of the Maxwell-J\"uttner distribution function (\ref{8}) into the definition of the energy-momentum tensor (\ref{5})$_2$ and the integration of the resulting equation, yielding
\be
e=mc^2\left[{K_3(\zeta)\over K_2(\zeta)}-{1\over\zeta}\right],\qquad p=nkT.
\ee{9}
Hence, the expression for the hydrostatic pressure is the same as the one for a non-relativistic gas. In the non-relativistic and ultra-relativistic limiting cases the energy per particle becomes
\be
e=mc^2+\frac{3}{2}kT\left[1+\frac{5}{4\zeta}+\dots\right],\qquad e=3kT+\dots\,,
\ee{9a}
respectively. Note that the first term in the non-relativistic limiting case (\ref{9a})$_1$ is the rest energy of a particle $mc^2$, the second one $3kT/2$ represents the classical value of the energy density, while the third one is a relativistic correction. By considering only the first two terms in this expansion it follows the well-known relationship between the pressure and the energy density of a non-relativistic gas, namely,  $p=2n(e-mc^2)/3$. Furthermore, in the ultra-relativistic limiting case  the following relationship $p=ne/3$ holds.

\subsection{Model equation and Chapman-Enskog method}

The  constitutive equations for the non-equilibrium pressure  $\varpi$, pressure deviator $\mathbb{P}^{\a\b}$ and heat flux $q^\a$ can be obtained from the Boltzmann equation by using the Chapman-Enskog methodology. Here we shall use the simple model equation proposed by Anderson and Witting \cite{AW}, which is a generalization of the BGK model for a relativistic gas. The model equation for the Boltzmann equation in this case is written as
\be
p^{\alpha}\left(U_\a Df+\nabla_\a f\right)=
-{U^{\alpha}p_{\alpha}\over c^2\tau}(f-f^{ (0)}),
\ee{10}
where $\tau$ is a characteristic time of order of the mean free time. Above we have introduced the differential operators $D$ and $\nabla_\a$ through the decomposition $\partial_\a=(U_\a D/c^2+\nabla_\a)$. The  differential operators $D$ and $\nabla_\a$ reduce to a partial time derivative and a spatial gradient in a Lorentz rest frame, respectively.

For processes close to equilibrium we may write $f=f^{(0)}(1+\phi)$, where the deviation $\phi$ from the Maxwell-J\"uttner distribution function  is considered as a small quantity. Now by using the Chapman-Enskog methodology we insert the equilibrium distribution function on the left-hand side of (\ref{10}) and the representation $f=f^{(0)}(1+\phi)$ into its right-hand side and get
\be
p^{\alpha}\left(U_\a Df^{(0)}+\nabla_\a f^{(0)}\right)=
-{U^{\alpha}p_{\alpha}\over c^2 \tau}\phi.
\ee{11}
Next we  perform the derivatives which occur in the left-hand side of above equation and eliminate the derivatives  $Dn$, $DT$ and $DU^\alpha$ by  using the balance equations for
the particle number density, energy density and momentum  density of an Eulerian relativistic gas (where  $\varpi=0$, $\mathbb{P}^{\a\b}=0$ and  $q^\a=0$), namely,
\be
Dn+n\nabla ^\alpha U_\alpha =0,\qquad
n c_v DT+p\nabla_\alpha U^\alpha=0,\qquad
{n\,h\over c^2}DU^\a=\nabla^\a p.
\ee{12}
Hence, the  deviation $\phi$ from the Maxwell-J\"uttner distribution function  can be determined, yielding
\ben\nonumber
&&\phi=-{c^2\tau\over U^{\alpha}p_{\alpha}}
\Biggl\{-{k^2T\over c^2c_v}\Biggl[
{\zeta^2c_v\over 3k}
-{1\over 3}\left(\zeta^2+5{K_3 \over K_2}\zeta-{K_3^2 \over K_2^2}\zeta^2-4\right)
\left({U^\beta p_\beta\over kT}\right)^2
\\\label{und}
&&-\left({K_3^2 \over K_2^2}\zeta^2-4{K_3 \over K_2}\zeta -\zeta^2\right)
\left({U^\beta p_\beta\over kT}\right)\Biggr]
\underline{\nabla _\alpha U^\alpha}-{p_\alpha p_\beta\over kT}
\underline{\widehat{\nabla^{ \alpha}U^{\beta}}}
+{p_\alpha\over  kT^2}(p_\beta U^\beta-h)\underline{\left(\nabla^\alpha T-
{T\over nh}\nabla ^\alpha p\right)}\Biggr\}.\qquad
\een
In the above equation we have introduced the   heat capacity per particle at constant volume $c_v$ and the traceless part of the velocity gradient $\widehat{\nabla^{\alpha}U^{\beta}}$ defined by
\be
c_v=\frac{\partial e}{\partial T}=k\Bigg(\zeta^2+5{K_3 \over K_2}\zeta-{K_3^2 \over K_2^2}\zeta^2-1\Bigg),\quad
\widehat{\nabla^{\alpha}U^{\beta}}=\Bigg[{\left(\Delta^{\alpha}_{\gamma}
\Delta^{\beta}_{\delta}
+\Delta^{\alpha}_{\delta}
\Delta^{\beta}_{\gamma}\right)\over 2}
-{ \Delta^{\alpha\beta}\Delta_{\gamma\delta}\over 3}
\Bigg]
\nabla^{\gamma}U^{\delta}.
\ee{14}
Note that  the deviation $\phi$ is a function of the thermodynamic forces which are underlined in (\ref{und}): the divergence of velocity $\nabla _\alpha U^\alpha$, the traceless part of the velocity gradient $\widehat{\nabla^{\alpha}U^{\beta}}$ and a combination of the temperature and pressure gradients $\left(\nabla^\alpha T-{T\over nh}\nabla ^\alpha p\right)$. The contribution of the pressure gradient refers only to a relativistic correction.

The  constitutive equations for the non-equilibrium pressure, pressure deviator and heat flux can be obtained now from the insertion of $f=f^{(0)}(1+\phi)$ together with (\ref{und}) into the definitions of the particle four-flow and energy-momentum tensor (\ref{5}) and integration of the resulting equations. Hence it follows
\be
\varpi=-\eta\nabla_{\alpha}U^{\alpha},\qquad
\mathbb{P}^{\a\b}=2\mu\widehat{\nabla^{ \alpha}U^{\beta}},
\qquad q^{\alpha}=\lambda\left(
\nabla^{\alpha}T-{T\over nh}
\nabla^{\alpha}p\right).
\ee{15}
The above equations represent the laws of Navier-Stokes and Fourier and we may identify  the scalar coefficients $\eta$, $\mu$ and $\lambda$  as the coefficients of bulk viscosity, shear viscosity and thermal conductivity, respectively. Their expressions read
\ben\label{15a}
\eta&=&{\tau p}\zeta\left[{{K_3^2 }\zeta-5{K_3  K_2}-\zeta K_2^2\over
\zeta^2K_2^2+5{K_3  K_2}\zeta-K_3^2 \zeta^2-K_2^2}+{\zeta^2\over 9}\left(
{3\over\zeta^2}{K_3 \over K_2}-{1\over\zeta}+{K_1\over K_2}-
{{\rm Ki}_1\over K_2}\right)\right],\qquad
\\\label{15b}
\mu&=&
{\tau p\over 15}\zeta^3\left[{3\over
\zeta^2}{K_3 \over K_2}-{1\over \zeta}+{K_1\over K_2}-
{{\rm Ki}_1\over K_2}\right],\qquad
 \lambda={\tau k p\over 3m}\zeta^4 {K_3 \over K_2}\left[{K_3 \over K_2}\left({1\over\zeta}
-{K_1\over K_2}+{{\rm Ki}_1\over K_2}\right)-{3\over \zeta^2}
\right],
\een
where ${\rm Ki}_1$ denotes the following integral for the modified
Bessel function ${\rm Ki}_1(\zeta)=
\int_0^\infty {e^{-\zeta\cosh t}\,dt/ \cosh t}$.

In the non-relativistic limiting case $(\zeta\gg1)$ the transport coefficients become
\ben
\eta={5\over 6}{p\tau\over \zeta^2}\left[1-{16\over \zeta}+\dots\right],\qquad
\lambda={5k\over 2m}p\tau\left[1-{3\over \zeta}+\dots\right],
\qquad
\mu=p\tau\left[1-{1\over \zeta}+\dots\right],
\een
while in the ultra-relativistic limiting case $(\zeta\ll1)$ they read
\ben
\eta={p\tau\over 54}\zeta^4
\left[1-{3\pi\over 2}\zeta+\dots\right],
\qquad
\lambda={4c^2\over 3T}p\tau\left[1-{11\over
8}\zeta^2+\dots\right],
\qquad
\mu={4\over 5}p\tau\left[1+{1\over
24}\zeta^2+\dots\right].
\een
In the non-relativistic limiting case the ratio $\lambda/\mu\approx5k/2m$ has the same value as that of the BGK model. Furthermore, we note that the bulk viscosity coefficient is important only in the relativistic case, since it vanishes in the non-relativistic and in the ultra-relativistic limiting cases.

Once  the constitutive equations (\ref{9}) and (\ref{15}) are known, the balance equations (\ref{6}) become a system of field equations for the five scalar fields of particle number density $n$, four-velocity $U^\a$ and temperature $T$ and problems concerning relativistic fluids in special relativity could be analyzed.

\section{Boltzmann equation in gravitational fields}

For the analysis of a relativistic gas in the presence of a gravitational field we have to consider a Riemannian space characterized by the metric tensor $g_{\mu\nu}$. Here the invariant elements in the phase space are $ {\sqrt{-g'}}{d^3p'/ p'_0}={\sqrt{-g}}{d^3p/ p_0}$ and ${g }d^3x d^3p/p_0={g'}d^3 x'd^3p'/ p'_0,$
 where $g=\det(g_{\mu\nu})$ is the determinant of the metric tensor. Moreover, the constant length of the
momentum four-vector $g_{\mu\nu}p^\mu p^\nu=m^2c^2$ implies that
\ben
p^0={p_0-g_{0i}p^i/ g_{00}}, \qquad\hbox{with}\qquad
p_0=\sqrt{g_{00}m^2c^2+\left(g_{0i}g_{0j}-g_{00}g_{ij}\right)p^ip^j}.
\een

The Boltzmann equation in gravitational fields is written as
\be
{ p^\mu{\partial f\over
\partial x^\mu}-{\Gamma_{\mu\nu}^i
p^\mu p^\nu{\partial f\over\partial p^i}}=
\int\left(f_*'f'-f_*f\right) \,F\,\sigma\,d\Omega\,
{\sqrt{-g}}\,{d^3p_*\over p_{*0}}},
\ee{18}
where $\Gamma_{\mu\nu}^\sigma=g^{\sigma\tau}\left(\partial_\mu g_{\mu\tau}+\partial_\nu g_{\mu\tau}-\partial_\tau g_{\mu\nu}\right)/2$ are Christoffel symbols.

Following the same methodology of the special relativistic case,  we may define  the two first moments of the distribution function and obtain their  respective balance equations from the Boltzmann equation (\ref{18}), which read
\ben\label{19}
&&N^\mu= c\int p^\mu f\,\sqrt{-g}\,{d^3p\over p_0},\qquad\qquad {N^\mu}_{;\mu}=\partial_\mu N^\mu+\Gamma_{\mu\sigma}^\mu N^\sigma= 0,\\\label{20}
&&T^{\mu\nu}= c\int p^\mu p^\nu f\,\sqrt{-g}\,{d^3p\over p_0},\qquad {T^{\mu\nu}}_{;\mu}=\partial_\mu T^{\mu\nu}+\Gamma_{\mu\sigma}^\mu T^{\sigma\nu}+\Gamma_{\mu\sigma}^\nu T^{\mu\sigma} =0.
\een
In the above equations the semicolon refers to a covariant derivative.

For the solution of a problem concerning a relativistic gas in the presence of a gravitational field, it is necessary to determine the ten components of the metric tensor $g_{\mu\nu}$, which is achieved by using Einstein's field equations
\be
R_{\mu\nu}-{1\over2}R g_{\mu\nu}=-\frac{8\pi G}{c^4}T_{\mu\nu},\qquad \hbox{where}\qquad R_{\mu\nu}=\partial_\nu\Gamma_{\mu\tau}^\tau-\partial_\tau\Gamma_{\mu\nu}^\tau+\Gamma_{\mu\tau}^\lambda\Gamma_{\nu\lambda}^\tau-\Gamma_{\mu\nu}^\lambda\Gamma_{\tau\lambda}^\tau,\qquad
R=g^{\mu\nu}R_{\mu\nu},
\ee{22}
represent the Ricci tensor and the curvature scalar, respectively. Moreover, $G$ denotes the gravitational constant. In the right-hand side of Einstein's field equations the energy-momentum tensor represents the sources of the gravitational field, while its left-hand side corresponds to the geometry of the curved space-time. The conservation of the energy-momentum tensor ${T^{\mu\nu}}_{;\mu}=0$ follows also from  Einstein's field equation thanks to the Bianchi identity $\left(R^{\mu\nu}-{1\over2}R g^{\mu\nu}\right)_{;\mu}=0.$

\subsection{Relativistic gas in a Robertson-Walker metric}

The search for cosmological solutions of our Universe are founded on Einstein's field equations and on the hypothesis that the Universe is spatially homogeneous and isotropic at large scales, i.e., of order of 100 Mpc $\approx 3.1\times10^{24}$ m $\approx 3.26\times10^8$ light-years. The isotropy and homogeneity assumptions imply that the line element $ds$ is given by the Friedmann-Lama\^{i}tre-Robertson-Walker (FLRW) metric, which for a spatially flat Universe reads
$ds^2=\left(c^2dt\right)^2- a(t)^2\left[\left(dx\right)^2+\left(dy\right)^2+\left(dz\right)^2\right].$
 This metric is a function of only one unknown parameter which is the cosmic scale factor  $a(t)$. In addition from the  hypothesis of homogeneity  and isotropy of the space-time, the pressure deviator $\mathbb{P}^{\a\b}$ and of the heat flux $q^\a$ must be absent from the energy-momentum tensor, and in this case  it reduces to ${T^\mu}_\nu={\rm diag}(en,-p-\varpi,-p-\varpi,-p-\varpi)$.

For the FLRW metric the non-vanishing components of the metric tensor and of the Christoffel symbols are given by
$g_{00}=1,$ $g_{ij}=- a ^2\delta_{ij},$  $g= -a^6,$ $\Gamma_{ij}^0=\dot a  a \,\delta_{ij},$ $\Gamma_{0j}^i=\delta^i_j\,{\dot a / a },$
where the dot denotes the derivative with respect to $x^0=ct$.

The components 00 and 11(say) of the  Einstein's field equations (\ref{22})$_1$ lead to  two  equations which may be used to analyze cosmological problems. These equations -- called the Friedmann and the acceleration equations --  read
 \be
\left({\dot a\over a}\right)^2={8\pi G \over3c^4}en,\qquad {\ddot a\over a}=-{4\pi G\over3c^4}\left[ en+3\left(p+\varpi\right)\right],
\ee{23}
respectively. In order  to obtain a closed system of equations for the cosmic scale factor $a$ and energy density $en$
it is necessary to close the system of equations (\ref{23}) with constitutive laws for the hydrostatic pressure $p$ and for the non-equilibrium pressure $\varpi$.
The thermal equation of state $p=p(en)$ gives the dependence of the pressure on the energy density. To obtain the constitutive equation for the non-equilibrium pressure,
we shall  use the Chapman-Enskog methodology for the Anderson and Witting model (\ref{18}). In the case of a homogeneous and isotropic Universe we may neglect the spatial gradients, so that  (\ref{18}) can be written as
\be
{\partial f\over\partial x^0}-2{\dot a \over a }p^i{\partial f\over
\partial p^i}=-{1\over c\tau}(f-f^{ (0)}),
\ee{25}
by considering a spatially flat Universe and a comoving frame where $(U^\mu)=(c, {\bf 0})$.
In the above equation the Maxwell-J\"uttner distribution function is given by
\be
f^{(0)}({\bf x},{\bf p},t)={n\over 4\pi m^2 c k T K_2(\zeta)} \exp\left[{-{c\sqrt{m^2c^2+a^2\vert{\bf p}\vert^2}\over k T}}\right].
\ee{26}
Note that $f^{(0)}$ depends on the cosmic scale factor $a$.

By following the same methodology above, we write $f=f^{(0)}(1+\phi)$ and insert this representation on the right-hand side of (\ref{25}) and on its left-hand side we perform the derivatives of  $f^{(0)}$. Hence, we get
\be
\phi=-c\tau\left[\frac{\dot n}{n}+\left(1-\zeta{K_3\over K_2}+{cp_0\over k T} \right)\frac{\dot T}{T}+{c\vert{\bf p}\vert^2 a\,\dot a
\over k T p_0}\right]=c\tau\left[3+\frac{3k}{c_v}\left(1-\zeta{K_3\over K_2}+{cp_0\over k T} \right)-{c\vert{\bf p}\vert^2 a^2
\over k T p_0}\right]{\dot a\over a}.
\ee{27}
The second equality above was obtained through the elimination of the time derivatives of the particle number density and of the temperature by considering the equations
\be
\dot n+3n{\dot a \over a}=0, \qquad \dot T+{3kT\over c_v}{\dot a \over a}=0,
\ee{28}
which follows from (\ref{19})$_2$ and from the 0th-component of (\ref{20})$_2$ with $\varpi=0$, respectively.

The following projection of the energy-momentum tensor
corresponds to the sum of the hydrostatic pressure
with the non-equilibrium pressure
\be
p+\varpi=-{c\over 3}\left(g_{\mu\nu}-\frac{U_\mu U_\nu}{c^2}\right)\int p^\mu p^\nu f\sqrt{-g}{d^3p\over p_0}
 ={c a ^5\over 3}\int\vert{\bf p}\vert^2\,f
{d^3p\over p_0}.
\ee{29}
In order to perform the above integral we introduce a new variable $y=p_0/mc=\sqrt{1+a^2\vert{\bf p}\vert^2/(mc)^2}$ so that we can write
\be
\vert {\bf p}\vert={m c \over a }(y^2-1)^{1\over 2},\qquad\hbox{and}\qquad
d\vert {\bf p}\vert={m c\over a }{ydy\over (y^2-1)^{1\over 2}}.
\ee{30}
Next, the insertion of $f=f^{(0)}(1+\phi)$ together with (\ref{27}) into
(\ref{29})  and by taking into account (\ref{30}), we get
\be
p+\varpi={p\zeta^2\over 3K_2}\int_1^\infty e^{-\zeta y}(y^2-1)
^{3\over 2}
\Biggl\{1+c\tau\left[3+\frac{3k}{c_v}\left(1-\zeta{K_3\over K_2}+\zeta y \right)-\zeta{y^2-1\over y}\right]{\dot a\over a}
\Biggr\}dy.
\ee{31}
From the  integration of the above equation we get the following
constitutive equation for the non-equilibrium pressure:
\be
\varpi=-3\eta
{\dot a \over a }.
\ee{32}
We identify $\eta$  in (\ref{32}) as the coefficient of bulk viscosity, since its expression is the same as that given by (\ref{15a}). Furthermore, from the comparison of (\ref{32}) with the constitutive equation for the non-equilibrium  pressure -- given in terms of the divergence of the
four-velocity (\ref{15})$_1$ --  we conclude that  $3 \dot a / a $ plays the same role as the divergence of the four-velocity. Owing to the fact that
 the bulk  viscosity is a positive quantity the non-equilibrium  pressure decreases when the Universe is expanding ($\dot a >0$) and it
increases when it is contracting ($\dot a <0$).

Now a cosmological problem can be solved from the system of coupled differential equations (\ref{23}) together with the constitutive equations for the pressure $p=p(en)$ and for the non-equilibrium pressure (\ref{32}), since in this case (\ref{23})  becomes a system of field equations for the determination of the cosmic scale factor $a(t)$ and energy density $ne(t)$.


\subsection{Relativistic gas in a Schwarzschild metric}

 Now we shall analyze a gas under the influence of a Schwarzschild metric, which is the solution of Einstein's field equation for  a  spherically symmetrical non-rotating and uncharged source of the gravitational field. In terms of the spherical coordinates $(\breve r, \theta, \varphi)$ the Schwarzschild metric reads
 \ben\label{33}
 ds^2={\left(1-\frac{2GM}{c^2 \breve r}\right)}\left(dx^0\right)^2-\left(1-\frac{2GM}{c^2 \breve r}\right)^{-1}d{\breve r}^2-{\breve r}^2\left(d\theta^2+\sin^2\theta d\varphi^2\right),
 \een
 where $M$ is the total mass of the spherical source. If we introduce a new radial coordinate $ r^2=\delta_{ij}x^ix^j$ through the relationship
  $ \breve r= r\left(1+\frac{GM}{2c^2 r}\right)^2$,
we get the isotropic  Schwarzschild metric in the  new radial coordinate
 \ben\label{34}
 ds^2=\frac{\left(1-\frac{GM}{2c^2 r}\right)^2}{\left(1+\frac{GM}{2c^2 r}\right)^2}\left(dx^0\right)^2-\left(1+\frac{GM}{2c^2 r}\right)^4\,\delta_{ij}\,dx^i \,dx^j
 \equiv g_0( r)\left(dx^0\right)^2-g_1( r)\,\delta_{ij}\,dx^i\, dx^j.
 \een

 For the isotropic metric (\ref{34}), the Christoffel symbols read
 \ben\label{34a}
 &&\Gamma_{00}^0=0,\qquad \Gamma_{ij}^0=0,\qquad \Gamma_{0j}^i=0,\qquad \Gamma_{ij}^k=0 \quad (i\neq j\neq k),\qquad
 \Gamma_{0i}^0=\frac{1}{2g_0( r)}\frac{d g_0( r)}{d r}\delta_{ij} \frac{x^j}{ r},
 \\\label{34b}
 &&\Gamma_{00}^i=\frac{1}{2g_1( r)}\frac{d g_0( r)}{d r}\frac{x^i}{ r},\qquad
 \Gamma_{\underline{i}\,\underline{i}}^j=-\frac{1}{2g_1( r)}\frac{d g_1( r)}{d r}\frac{x^j}{ r}\quad (i\neq j),\qquad
 \Gamma_{\underline{i}\,j}^{\underline{i}}=\frac{1}{2g_1( r)}\frac{d g_1( r)}{d r}\delta_{jk} \frac{x^k}{ r},
 \een
 where the underlined indices are not summed. Furthermore, it follows that $p_0=\sqrt{g_0}\sqrt{m^2c^2+g_1\vert {\bf p}\vert^2}$, $p^0={p_0}/{g_0}$, $\sqrt{-g}=\sqrt{g_0g_1^3}$,
while the components of the four-velocity in a comoving frame reads $U^\mu=\left({c}/{\sqrt{g_0}},\bf{0}\right)$.

Now by writing  $f=f^{(0)}(1+\phi)$ and following the Chapman-Enskog methodology we get from  the Boltzmann equation (\ref{18})
\ben\nonumber
&&f^{(0)}\Bigg\{\frac{p^\nu}{n}\frac{\partial n}{\partial x^\nu}+\frac{p^\nu}{T}\left[1-\frac{K_3}{K_2}\zeta +\frac{ p^\tau U_\tau}{kT}\right]\frac{\partial T}{\partial x^\nu}-\frac{p^\tau p^\nu}{kT}\left(g_{\tau\epsilon}\,{U^\epsilon}_{;\nu}+\Gamma_{\tau\nu}^\sigma\, g_{\sigma\epsilon}\,U^\epsilon\right)+\frac{g_0U^0}{k T}g_1\delta_{ij}\frac{p^j}{p_0}\Gamma_{\sigma\nu}^ip^\sigma p^\nu \\\label{35}
&&+\frac{g_0U^0}{kT}p^\nu\left(\frac{p^0}{2g_0}\frac{\partial g_0}{\partial x^\nu}-\frac{1}{2p_0}\frac{\partial g_1}{\partial x^\nu}\delta_{ij}p^ip^j\right)
\Bigg\}=\int f^{(0)}f_*^{(0)}\left(\phi_*'+\phi'-\phi_*-\phi\right) \,F\,\sigma\,d\Omega\,
{\sqrt{-g}}\,{d^3p_*\over p_{*0}}.
\een

The multiplication of (\ref{35}) by $\sqrt{-g}d^3 p/p_0$ and integration of the resulting equation leads to the balance equation of particle number density
\be
U^\nu\,\frac{\partial n}{\partial x^\nu}+n\, {U^\nu}_{;\nu}=0.
\ee{35a}
Furthermore, the multiplication of (\ref{35}) by $p^\mu\sqrt{-g}d^3 p/p_0$ and subsequent integration  imply into the balance equations for the energy density and momentum density of an Eulerian fluid. They read
\ben\label{35b}
&&n \,c_v\,U^\nu\,\frac{\partial T}{\partial x^\nu}+p\, {U^\nu}_{;\nu}=0, \qquad 0-\hbox{th component},
\\\label{35c}
&&n\,m\frac{K_3}{K_2} U^\nu\,{U^k}_{;\nu}-g^{ki}\frac{\partial p}{\partial x^i}
-\underline{n\,m\,c^2\frac{K_3}{K_2} g^{ki}\frac{1}{g_0}\frac{d g_0}{dr}\delta_{ij}\frac{x^j}{r}}
=0,\qquad k-\hbox{th component},
\een
respectively.

It is interesting to  discuss the balance equation for the momentum density (\ref{35c}), since  the underlined term refers to the presence of the gravitational field.
By introducing the gravitational potential
\ben\label{36}
\Phi=-\frac{GM}{r},\qquad\hbox{so that} \qquad \frac{\partial \Phi}{\partial x^k}=\frac{GM}{r^2}\delta_{kj}\frac{x^j}{r},
\een
we can write the balance equation for the momentum density (\ref{35c}) as
\ben\label{37}
m\,n\,\frac{K_3}{K_2}U^\mu {U^k}_{;\mu}-g^{ki}\frac{\partial p}{\partial x^i}
-m\,n\,\frac{K_3}{K_2}\frac{1}{1-\Phi^2/4c^4}g^{k i}\frac{\partial\Phi}{\partial x^i}=0,
\een

The ratio $\Phi/c^2$ can be estimated at the surface of some bodies:
\begin{enumerate}
\item  Earth: $M_{\oplus}\approx 5.97\times
10^{24}$ kg; $R_\oplus\approx 6.38\times 10^6$ m;
${\Phi/ c^2}\approx 7 \times 10^{-10}$;
\item   Sun: $M_{\odot}\approx 1.99\times
10^{30}$ kg; $R_\odot\approx 6.96\times 10^8$ m; ${\Phi/
c^2}\approx 2.2 \times 10^{-6}$;
\item   White dwarf: $M\approx 1.02 M_{\odot}$;
$R\approx 5.4\times 10^6$ m;
${\Phi/ c^2}\approx 2.8 \times 10^{-4}$;
\item   Neutron  star: $M\approx
M_{\odot}$; $R\approx 2\times 10^4$ m;
${\Phi/ c^2}\approx 7.5 \times 10^{-2}$;
\item   Black hole: $M\approx 3 M_{\odot}$;
$R\approx 3\times 10^3$ m;
${\Phi/ c^2}\approx 1.5$.
\end{enumerate}
From the above estimates we infer that in most cases we can use the approximation $\Phi/c^2\ll1$.

Hence, when $\Phi/c^2\ll1$ and by considering  the non-relativistic limiting case $\zeta\gg1$, the balance equation of the momentum density (\ref{37}) becomes
\ben
m\,n\,\left(1+\underline{\frac{5}{2\z}+\dots}\right)U^\mu {U^\nu}_{;\mu}-\Delta^{\mu\nu}\frac{\partial p}{\partial x^\mu}
-m\,n\,\left(1+\underline{\frac{5}{2\z}+\dots}\right)\left(1+\underline{\frac{\Phi^2}{4c^4}+\dots}\right)g^{\nu i}\frac{\partial\Phi}{\partial x^i}=0.
\een
By neglecting  the underlined terms the above equation reduces to the usual Newton's second law in the case of a  non-relativistic gas in the presence of a weak gravitational field.
The underlined terms represent the relativistic and the gravitational potential corrections.

\section{Conclusions}

In this work we have derived the field equations for relativistic gases in special and general relativity from the Boltzmann equation. The transport coefficients of a relativistic gas in special relativity were obtained from the Anderson and Witting model equation. In general relativity we have analyzed first a homogeneous, isotropic and flat Universe, where irreversible processes are taken into account and derived the corresponding field equations. Next, a relativistic gas under the influence of  a  spherically symmetrical non-rotating and uncharged source of the gravitational field was considered and the corresponding balance equations were obtained.

\begin{theacknowledgments}
 The paper was partially supported by Brazilian Research Council (CNPq).
 \end{theacknowledgments}


 \bibliographystyle{aipproc}

\begin{thebibliography}{9}

 \bibitem{J1}
F. J\"uttner, 
{\it Ann. Physik und Chemie}  {\bf 34}, 856-882 (1911).

\bibitem{J2}
F. J\"uttner, 
{\it Zeitschr. Physik} {\bf 47}, 542-566 (1928).

\bibitem{LM} A. Lichnerowicz and R. Marrot, \emph{C. R. Acad. Sci.} \textbf{210}, 759-761 (1940).

\bibitem{I} W. Israel,  \emph{J. Math. Phys.}  \textbf{4}, 1163-1181 (1963).

\bibitem{K} D. C. Kelly, {\it The kinetic theory of a relativistic gas},
unpublished report (Miami University, Oxford, 1963).

  \bibitem{C} N. A. Chernikov, \emph{Acta Phys. Polon.} \textbf{23}, 629-645 (1963).

 \bibitem{AW} J. L. Anderson and  H. R. Witting, 
 \emph{Physica} \textbf{74}, 466-488 (1974).

 \bibitem{1}
 C. Cercignani and G. M. Kremer, \emph{The relativistic Boltzmann equation: theory and applications} (Birkh\"auser, Basel, 2002).


 \bibitem{2}
 G. M. Kremer, \emph{Relativistic fluids in special and general relativity} in Proceedings of the IV Mexican Meeting on Mathematical and Experimental Physics: Relativistic Fluids and Biological Physics, ed. L. Dagdug, A. L. Garc\'{\i}a-Perciante, A. Sandoval-Villalbazo and L. S. Garc\'{\i}a-Col\'{\i}n, pp. 3-13, AIP Proceedings 1312, Melville 2010.


\end{thebibliography}

\end{document}